# A Survey on Delay-Aware Network Structure for Wireless Sensor Networks with Consecutive Data Collection Processes

Ms.Aruna.G.R[1], Mr.SivanArulSelvan[2].

*1-Student, Department of CSE, Kumaraguru college of Technology, Coimbatore-49.*

*2-Associate Professor, Department of CSE, Kumaraguru college of Technology, Coimbatore-49.*

*TamilNadu, India.*

*Abstract—* **A Wireless Sensor Network (WSN) consists of spatially distributed autonomous sensors to monitor physical or environmental conditions, such as temperature, sound, pressure, etc. In sensing applications, data packets are flowing from sensor nodes to base station. In data collection processes, bottom up approach is used. In bottom up approach, all nodes send their sensed data packets to base station directly. In this approach will lead to increased delay, which will lead to higher energy consumption. To reduce the energy consumption of sensor nodes, Clustering Algorithm and Low-Energy Adaptive Clustering Hierarchy (LEACH) are being used. Efficient gathering in Wireless Sensor information systems, Power Efficient Gathering in Sensor Information System (PEGASIS) is being used. There are lot of research issues in Wireless Sensor Networks such as delay, lifetime of network, energy dissipation which needs to be resolved.**

## I. INTRODUCTION

Wireless Sensor Networks (WSN) consists of highly distributed networks deployed in large numbers to monitor the environment or systems by the measurement of physical parameters such as temperature, pressure or relative humidity. Each node in the sensor network consists of three subsystems namely sensor subsystem, processing subsystem and communication subsystem. The sensor subsystem is used sense the environment. The processing subsystem performs local computation on the sensed data. The communication subsystem is responsible for message exchange with neighbouring sensor nodes.

Two types of WSN are Unstructured WSN and Structured WSN. Unstructured WSN consists of dense collection of nodes and there is difficulty in network maintenance. Structured WSN consists of few and rarely distributed nodes and there is lower network maintenance. Sensor nodes are used in variety of application which requires constant monitoring and specific event detection. In military application, sensor node includes battlefield surveillance and monitoring. Environmental application includes detection of forest fire, flood and habitat exploration of animals. Sensor nodes can also be used in patient diagnosis and day to day home appliances.

Architecture of WSN includes Layered Architecture and Clustered Architecture. In Layered Architecture have layers of sensor nodes around the single powerful Base Station (BS). In Clustered Architecture sensor nodes are grouped into a cluster. Each cluster has a Cluster Head (CH) [1].Sensor nodes communicate with the Cluster Head and the Cluster Head communicates with the BS. Some of the issues of WSN are energy efficiency, delay, limited storage and computation, high error rate.

Delay is one critical issue in WSN which causes energy consumption. Delay is nothing but time taken by receiver to receiving a packet is higher than the time taken by the sender to send a packet. So delay causes high energy consumption. Low Energy Adaptive Clustering Hierarchy (LEACH) protocol is used to reduce energy consumption. In LEACH protocol, self organization of sensor nodes will occur [4].

## II. DATA COLLECTION PROCESS

Wireless Sensor Networks (WSNs) consists of large number of sensor nodes which can be used in any environment. Sensor nodes are compact, light-weighted and battery powered devices which can be used in any environment. In order to do close-range sensing, sensor nodes are deployed nearer to target. Collection of information from the sensor nodes is called Data Collection Process. The collected data will be sent to users who are located in a remote site. If the sensor nodes are located in extreme environment, maintenance will be difficult. Clustering is effective for Data Collection Process. Consecutive Data Collection Process is successive collection of information from the sensor nodes. It is collected from sensor nodes periodically. In consecutive Data Collection Process, clustering is not effective.

*A. Clustering Algorithm*

Conservation of energy is done by clustering. A network with large number of sensor nodes is divided into several clusters. Within each cluster, one of the sensor node is elected as Cluster Head(CH).Nodes other than Cluster Head(CH) are Cluster Members(CM).The Cluster Member can communicate with the cluster head and the cluster head can communicate with the remote Base station(BS).Collection of data from cluster member to cluster head can be done in multi-hop manner. The cluster head has several responsibilities such as collection of data from the cluster members, fusing those data by data fusion techniques and reporting those fused data to remote Base station [2]. In long distance transmission, number of nodes involved will be decreased, so that energy dissipation will also be reduced [8]. Delay will be decreased in clustering.

Disadvantage of clustering is energy draining of the cluster head will occur because all nodes in the cluster communicates with the cluster head and the cluster head communicates with the remote base station. This will affect the entire cluster hierarchy [2].

### B. LEACH Protocol

LEACH is cluster formation technique which enables self-organization of large number of nodes. It has algorithms which are used to adapt the clusters and to rotate the cluster head which will be useful to distribute the energy load among all the nodes. LEACH design goals are: a) randomized, adaptive, self-configuring cluster formation, b) localized control for data transfer, c) low-energy media access control, and d) application-specific data processing [3].

Normally many nodes are available in wireless sensor networks. In LEACH, self configuration of nodes into local clusters occurs, and then cluster head will be elected from those nodes. Nodes other than cluster head transmit their data to cluster head. Cluster head performs signal processing functions on data and transmit those data to the remote base station. Therefore, cluster head node is more intensive than non-cluster head node. Cluster head position will be rotated evenly among sensors to avoid the draining of battery in any one of the sensor in the network. The LEACH operation's can be divided into rounds. During cluster organization, each round starts with set-up phase followed by steady-state phase when the data are transferred from the nodes to cluster head and on to the base station. Fig 1. shows the cluster head selection, and distributed cluster formation algorithms and the steady-state operation of LEACH.

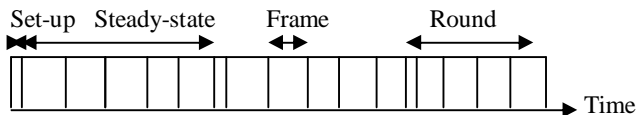

Fig 1. Time line showing LEACH operation.

*1) Cluster Head Selection Algorithm:*

LEACH uses distributed algorithm to form clusters. In distributed algorithm, nodes make independent decisions without any centralized control. There are number of clusters k during each round, the goal is to form cluster formation algorithm in those clusters. If the nodes begin with equal energy, the goal is to evenly distribute the energy load among nodes in the network. It will avoid overly utilized nodes. Each node acts as a cluster head because cluster head is more intensive than cluster [3]. The probabilities of cluster head selection are shown in (1) and (2)

$$P_i(t) = \begin{cases} k/N-k*(r \bmod N/k) & C_i(t)=1 \\ 0 & C_i(t)=0 \end{cases} \quad (1)$$

$$P_i(t) = \min\{E_i(t)/E_i(total)\} \quad (2)$$

*2) Cluster Formation Algorithm:*

Advertisement message (ADV) is broadcasted using a non persistent carrier-sense multiple access (CSMA) MAC protocol. It is a small message which contains node's ID and a header. A header is used to distinguish the message as an announcement message. Each non-cluster head node selects the cluster head. Selection is based on the received signal strength of the advertisement. For pure signal strength, advertisement of cluster head with largest signal strength becomes cluster head. It will require minimum energy for communication [7]. Random cluster head will be selected if there is an obstacle impending communication. After becoming a member of the cluster, it must inform the cluster head node that it is a member of the cluster. Join-request (Join-Req) message is transmitted by each node using a non-persistent CSMA MAC protocol.

The cluster head node with TDMA schedule will avoid collision among data messages. It also allows transmitting data during active state. After this, steady-state operation begins. The message is a short message. It consists of the node's ID and the cluster head's ID [3].

In LEACH, the cluster head act as a local control centre. It is used to coordinate the data transmission in the cluster. Each cluster head sets up a TDMA schedule that can be transmitted to the nodes in the cluster.

*3) Steady State Phase:*

The steady state operation is divided into frames. During the allocated transmission slot, nodes can transmit the data to the cluster head at most once per frame. In the duration of each slot, each node transmits the data to the cluster head in constant rate. So the time to send a frame of data depends on the number of nodes in the cluster.

To reduce energy dissipation, based on the received strength of the cluster head advertisement, each non cluster head node uses power control to set the amount of transmits power. The cluster head must be awake to receive all the data from the nodes. To enhance the common signal and reduce the uncorrelated noise, the data aggregation can be done after the cluster head receives all the data from the nodes in the cluster.

Perfect correlation can be assumed such that all individual signals can be combined into a single representative signal [6]. The finalized data are sent to the base station from the cluster head. If the base station is far away and the data messages are large, then the transmission is said to be high energy transmission.

The MAC and routing protocol are designed to ensure low energy dissipation in the nodes. It also ensures no collision of data messages within a cluster. Radio is inherently a broadcast medium, so that the transmission gets affected if the two clusters are neighbour to each other. Each cluster in LEACH communicates using Direct Sequences Spread

Spectrum (DSSS), in order to reduce the inter-cluster interference each cluster uses a unique spreading code.

By using this spreading code, all the nodes in the cluster transmit the data to the cluster head that filters all received energy [7]. This is known as transmitter based code assignment.

*C. PEGASIS*

In order to reduce the energy consumption in local communication, organize the nodes into a single chain. To reduce the load of the cluster head, distribute the load to all the nodes in the network by changing the cluster head from time to time. In chain formation, the greedy approach cannot guarantee a constant separation of the adjacent nodes [4].

The main objective of PEGASIS is for each node can transmit to and receive from the neighbour nodes and takes a turn being the leader for the transmission to the BS. This approach mainly helps to distribute the load among the sensor nodes in the network

The nodes can be formed into chain. The chain formation can be done by either be computed in a centralized manner by the BS, broadcast the data to all nodes or by using the greedy algorithm to accomplish the sensor nodes by themselves.

If the chain is computed by the sensor nodes, the nodes can first get the location information of the sensors locally and compute the chain by using the same greedy algorithm. All nodes having the same location data and run the same greedy algorithm then it produces the same result.

The energy cost for the data communication overhead is small compared to the energy spent in the data collection phase. The energy cost for the data collection, fusion and the transmission to the BS should be considered and evaluate it after the first node dies [4].

Assume there is no mobility; it leads to there will be no change in the chain in case of PEGASIS. To construct the chain, assume that all nodes have the global knowledge about the network and then employ the greedy algorithm.

Before the first round of communication, the greedy approach to constructing the chain works well. To construct the chain, first select the furthest node from the BS. The next node in the chain must be the closest neighbour node to that furthest node. To form the greedy chain, select the successive neighbours in this manner among the unvisited nodes.

The farthest node should be selected first in order to confirm that nodes farther from the BS have close neighbours as, in greedy algorithm, the distances of the neighbours will gradually increases since nodes already on the chain cannot be revisited.

Each round of the data collection process can be initiated by the BS. The beacon signal will synchronize all sensor nodes to transmit the data [4]. Since the positions of all the sensors are known, the time slot approach is employed to transmit the data.

A simple control token passing approach is initiated by the leader. The leader starts the data transmission from the ends of the chain. If the token size is small, then the cost is also small.

Every node performs data fusion except the nodes in end of the chain. Each node will fuse the neighbour data by its own. The node generates the fused data into a single packet of same length [6]. Then, transmit that packet to its other neighbour node. PEGASIS protocol is more advanced energy saving than LEACH protocol.

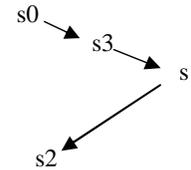

Fig 2.Chain construction using greedy algorithm.

Fig 2. represents node s0 connecting to node s1, then node s1 connects to node s2, and then node s2 connects to node s3, in that order. The chain will be reconstructed in the same manner to bypass the dead node when node dies.

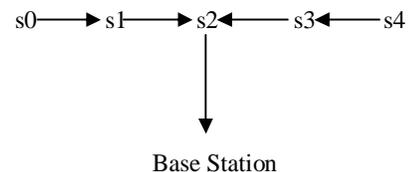

Base Station
Fig 3.Token Passing Approach

A simple token passing approach is used. In that approach, leader starts the data transmission from the ends of the chain. Token size is very small, so the cost is also very small. In Fig 3. node s2 is the leader. It will pass the token first to node s0 along the chain. The data is transmitted to node s2 from s0. After the node s2 receives data from node s1, it will pass the token to node s4. Node s4 will pass its data towards node s2.Here data fusion taking place along the chain.

PEGASIS able to perform data fusion takes place at every node except the end nodes in the chain. A single packet of same length can be generated by fusion of neighbour's data with its own which was done by each node. In PEGASIS, except the two end nodes and, the leader node, each node receives and transmits one data packet in each round. It will be the leader once every N rounds. A very small control token packet are received and transmitted by the nodes.

In different size networks and random node placements, greedy chain construction performs well. Some nodes may have relatively distant neighbours along the chain during the construction of chain. Such nodes will dissipate more energy when compared to other sensors. The performance of PEGASIS is improved by not allowing such nodes to become leaders. By setting threshold value on neighbour node to be leaders, performance of PEGASIS is improved slightly. The chain will be reconstructed and the threshold can be changed whenever node dies which is used to determine which nodes to be leader.

TABLE I
COMPARSION BETWEEN CLUSTERING ALGORTITHM, LEACH PROTOCOL, PEGASIS

|  | Energy X delay cost | | | | |
| --- | --- | --- | --- | --- | --- |
|  | Energy | | Delay | Energy $^x$ Delay | |
|  | D=50 | D=100 | D=50& D=100 | D=50 | D=100 |
| Clustering Algorithm | 0.3278 | 1.2801 | 100 | 32.7838 | 128.0159 |
| LEACH | 0.0788 | 0.2035 | 36 | 2.8368 | 7.3260 |
| PEGASIS | 0.0238 | 0.0355 | 100 | 2.3808 | 3.5507 |

The above table shows the comparison between Clustering Algorithm, LEACH Protocol, and PEGASIS in terms of Energy X delay cost. The above three techniques are compared by selecting random 50 and 100-node networks. For calculating Energy X delay, multiply the average energy cost per round to the unit delay. Dividing the nodes into 10 groups give the optimal Energy X delay for 50-node, 100- node, 200-node networks.

## III. ISSUES

There are some other issues in Wireless Sensor Networks such as formation of clusters is difficult [5]. High performance of LEACH can be achieved in tight constraints [3]. Higher order dissipation models are difficult to achieve in PEGASIS [4].

Energy saving is crucial for slow and infrequent target tracking applications [7]. Responsiveness can be achieved by exchanging more control packets. It will result in less reliability and less scalability [9]. In Wireless Sensor Networks, if the nodes stay longer to save power, it will lead to less communication with its neighbour node. Packet loss and congestion occurs due to limited Bandwidth.

In network design point of view, apriori knowledge about deployment of nodes is difficult [10]. To increase coverage, cluster or tree topology is implemented. In this architecture, each node maintains a single communication path with the gateway. Here all nodes depending upon the route node will lose their communication paths to the gateway when the router nodes go down.

## IV. CONCLUSIONS

LEACH saves energy when compared with clustering algorithm. In LEACH, computation is done locally in order to reduce the amount of transmitted data and MAC and routing protocols enable low-energy networking. Under tight constraints of the wireless channel, LEACH provides the high performance [3]. PEGASIS is greedy chain protocol. Since PEGASIS eliminates the overhead of dynamic formation, PEGASIS works better than LEACH [4]. In order to avoid the node death at random location, nodes takes turn to transmit the fused data to the BS. This helps to balance the energy depletion in the network. LEACH is decentralized algorithm. It forms multiple clusters with two-hop topology. PEGASIS is centralized algorithm. Sensor nodes form a single chain after sensor nodes are sorted.